\documentstyle[11pt,paspconf]{article}
\input{psfig}
\begin{document}

\title{Enabling Concepts for a Dual Spacecraft 
Formation-Flying Optical Interferometer
for NASA's ST3 Mission}
\author{P. W. Gorham, W. M. Folkner, and G. H. Blackwood}
\affil{Jet Propulsion Laboratory, California Institute of Technology,
4800 Oak Grove Drive, Pasadena, CA, 91109}

\begin{abstract}
We present the enabling concept and technology for
a dual spacecraft formation-flying optical
interferometer, to be launched into a deep space
orbit as Space Technology 3. The
combiner spacecraft makes use of a nested cat's
eye delay line configuration that minimizes
wavefront distortion and stores 20 m of optical
pathlength in a package of $\sim 1.5$ m length. A
parabolic trajectory for the secondary
collector spacecraft enables baselines of
up to 200 m for a fixed 20 m stored delay and 
spacecraft separations of up to 1 km.
\end{abstract}


\section{Introduction}

The last 15 years has seen considerable progress in the conception
and development of ideas for multi-spacecraft optical interferometers.
Stachnik, Melroy, and Arnold (1984) first laid out the conceptual
framework for an orbiting Michelson interferometer, and the following
year the European Space Agency devoted a colloquium to spacecraft
arrays of this type. Particular emphasis in a number of studies 
(Johnston \& Nock 1990; DeCou 1991) was placed on the choice of
orbits to minimize fuel usage and provide maximal UV coverage. 
DeCou (1991) described a family of orbits near
geostationary which were particularly efficient in this respect.

In the early 1990's a coherent effort began at the
NASA/Caltech Jet Propulsion Laboratory to
develop a consistent and detailed design for a three-spacecraft Michelson
interferometer known initially as the Separated Spacecraft Interferometer
(SSI; Kulkarni 1994), and then as the New Millenium Interferometer (NMI;
McGuire \& Colavita 1996; Blackwood et al 1998) because of its alignment
with NASA's New Millenium Technology program, in which it was scheduled
as Deep Space 3 (DS3).

However, funding constraints
eliminated the original three-spacecraft baseline design in late 1998 and a
de-scoped version involving somewhat reduced capability and requiring
only two spacecraft, was adopted.
The new mission, known as Space Technology 3 (ST3) due to
re-alignment of the parent NASA program, has moved into a
prototype construction phase, with launch now set for 2005.

In this report we describe the primary enabling concepts for 
the dual spacecraft system, which combine specific choices of
array geometry with a novel fixed optical delay line capable of supporting
a continuously variable interferometer baseline from 40 to 200 m.
The next section describes the choice of geometry, followed by a section
describing the overall optical layout, and the fixed delay line.
A related paper in this volume (Lay et al.) describes in detail the
operation of the interferometer system. 

\section{Observing Geometry}

\subsection{Original dual-spacecraft concept}

\begin{figure}[t]
\psfig{file=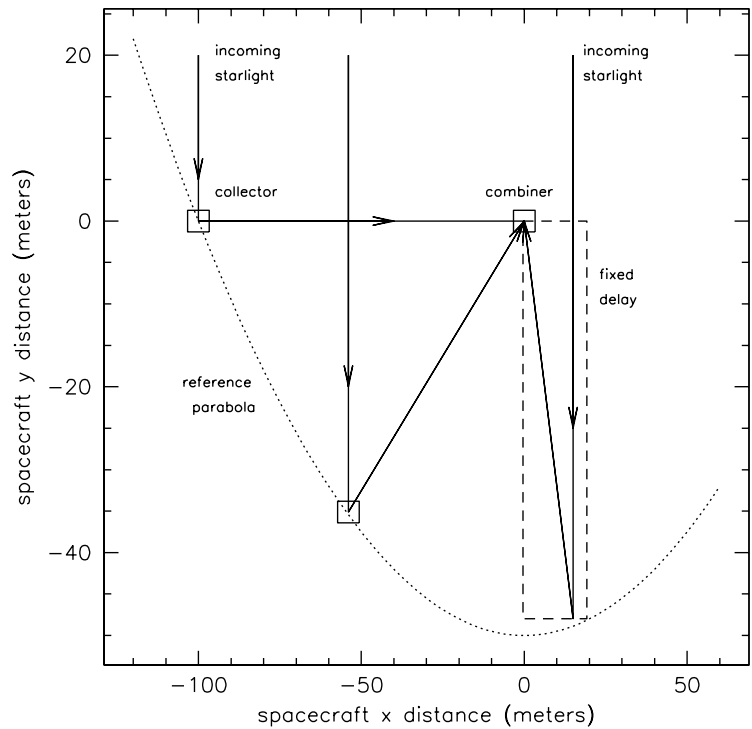,width=4.1in}
\caption{Original 2-spacecraft geometry (Folkner 1996). Note
different x and y scales.} \label{fig-1}
\end{figure}

Before the initial three-spacecraft configuration for DS3 was adopted
as the working design, a variety of different configurations were
considered which gave various levels of technology demonstration
with respect to a formation--flying multiple spacecraft interferometer.
One of these proposed early configurations was in fact a dual--spacecraft
system (Folkner 1996). The basic geometry of this configuration is
shown in Figure 1. Here the collector spacecraft (which acts simply
as a moving relay mirror) travels along a parabolic trajectory with
the combiner spacecraft at the focus of the parabola, which we choose
as the origin in this plot. The combiner spacecraft then carries a
fixed optical delay line which compensates for the additional
pathlength that the collector spacecraft produces. This is indicated
schematically by showing the fixed delay line as if it were
reflecting off another relay mirror at the surface of the
reference parabola, thus ensuring equal delay in the two arms of
the interferometer.

\begin{figure}[t]
\psfig{file=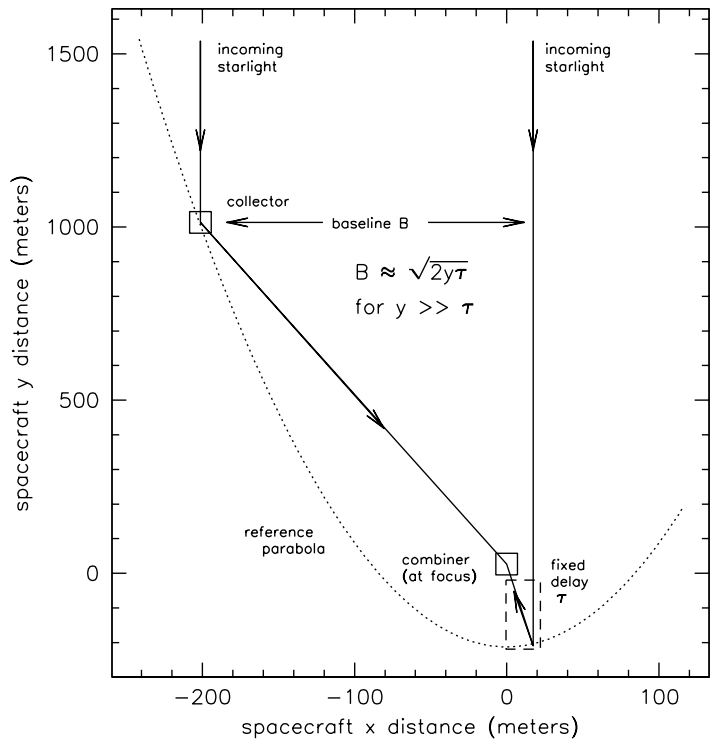,width=4.1in}
\caption{Geometry adopted for ST3, using opper portion
of reference parabolic trajectory. Note differing x and y scales.} \label{fig-2}
\end{figure}

For the pictured geometry, the collector spacecraft position $(x,y)$
must satisfy:
\begin{equation}
\sqrt{B^2 ~+~ y^2} ~=~ \tau ~+~ y
\end{equation}
where $x$ coordinate is defined as the projected baseline $B$, and the
total fixed delay carried by the combiner spacecraft is $\tau$. In
the case of Fig. 1 the $y$-position of the collector spacecraft was 
always negative with respect to the combiner for simplicity in the
relay optics. Equation (1) then determines the required collector
spacecraft position for a given projected baseline
\begin{equation}
y ~=~ {B^2 \over 2 \tau} \left [ 1 - {\tau^2 \over B^2} \right ]~.
\end{equation}
For the configuration of Fig. 1, the fixed delay is $\tau=100$ m,
and the maximum baseline (at $y=0$) is then also 100 m. 

The difficulty with this approach is the requirement that the
combiner spacecraft must carry a 100 m fixed delay line in a very compact
configuration, of order 1-2 m in overall length. This amount of delay
is not easily achievable in a broad-band system (450-1000 nm) as was planned 
for DS3. Approaches involving $\sim 50$ reflections between opposing 
spherical or flat mirrors typically produce too much wavefront
distortion, absorption, and scattering losses to be useful for 
a white light interferometer. Alternatives such as the use of
optical fiber also do not afford the broadband single-mode operation
required for a delay line.

\subsection{Modified approach}

Figure 2 shows a modified approach to the two spacecraft system in
which a much shorter fixed delay line can be utilized. Here the
spacecraft configuration entails a collector spacecraft position
which moves along the reference parabola {\em above} the combiner
spacecraft with respect to the source direction. Referring to
equation (2) above, $y=0$ when $B=\tau$. When $B$ exceeds the fixed delay
$\tau$, the collector spacecraft $y$-value then becomes positive.
For $B>>\tau$, 
\begin{equation}
y ~\approx~ {B^2 \over 2 \tau}~;
\end{equation}
thus the interspacecraft distance $D=\sqrt{y^2+B^2}$ grows 
quadratically with baseline. 

For the de-scoped NASA mission ST3, preliminary design considerations
indicated that a fixed delay line of $\sim 20$ m stored delay was
achievable within the constraints of spacecraft size and
instrument visibility budget. In a later section we provide details of
the fixed delay line design. Using $\tau = 20$ m, and the additional
constraint of $D\leq 1$ km imposed by formation--flying requirements,
ST3 is able to achieve a maximum interferometer baseline of about 200 m.

\section{Optical Design}

\begin{figure}
\psfig{file=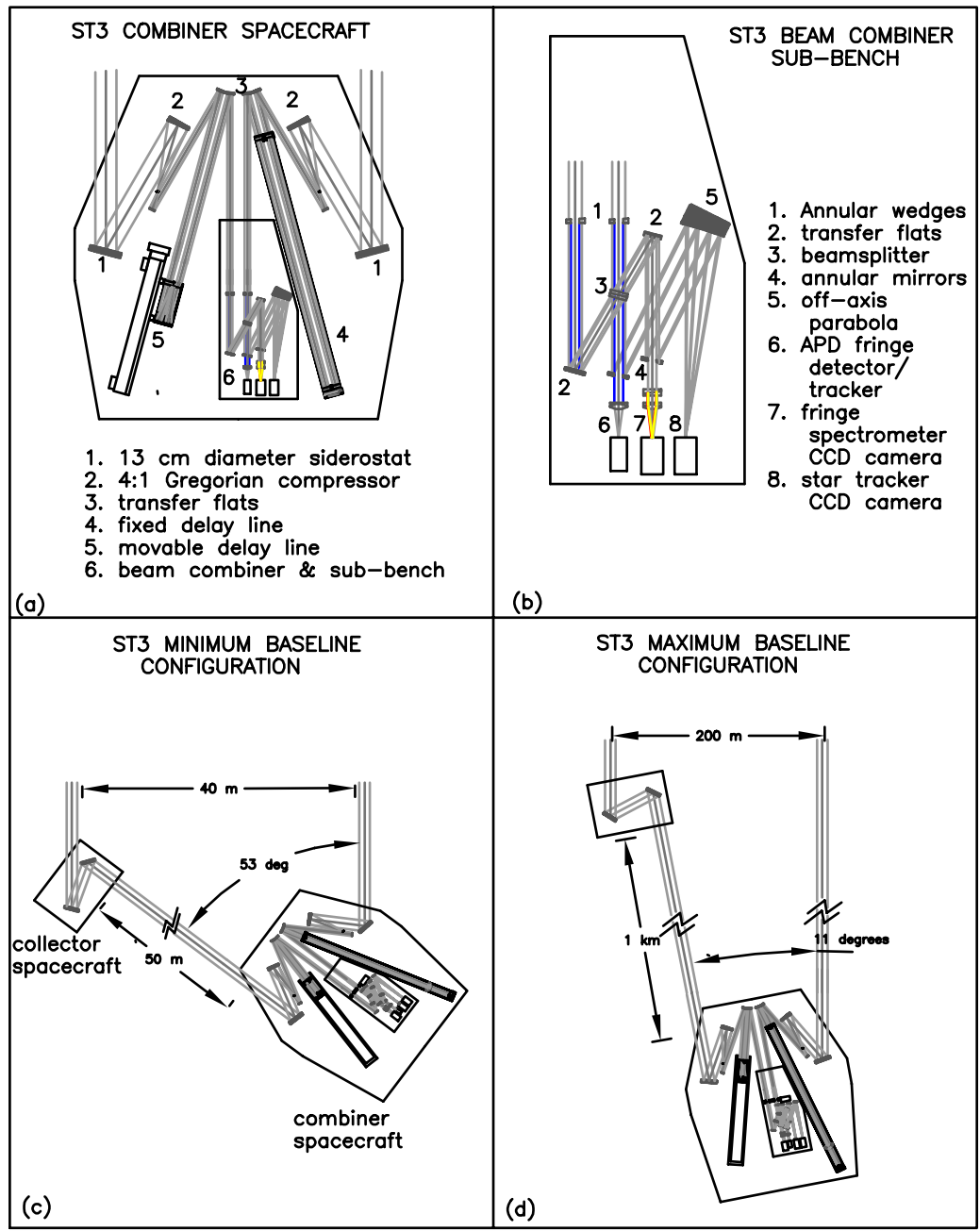,width=5.4in}
\caption{(a) Schematic optical layout of the combiner spacecraft
interferometer instrument for ST3. (b) A detail of the beam-combiner
sub-bench optical layout. (c) Schematic configuration geometry for
ST3 minimum baseline, (40 m; $D=50$ m). (d) Schematic configuration in the
maximum (200 m) baseline configuration, with 1 km spacecraft
separation.} \label{fig-3}
\end{figure}

\begin{figure}[t]
\psfig{file=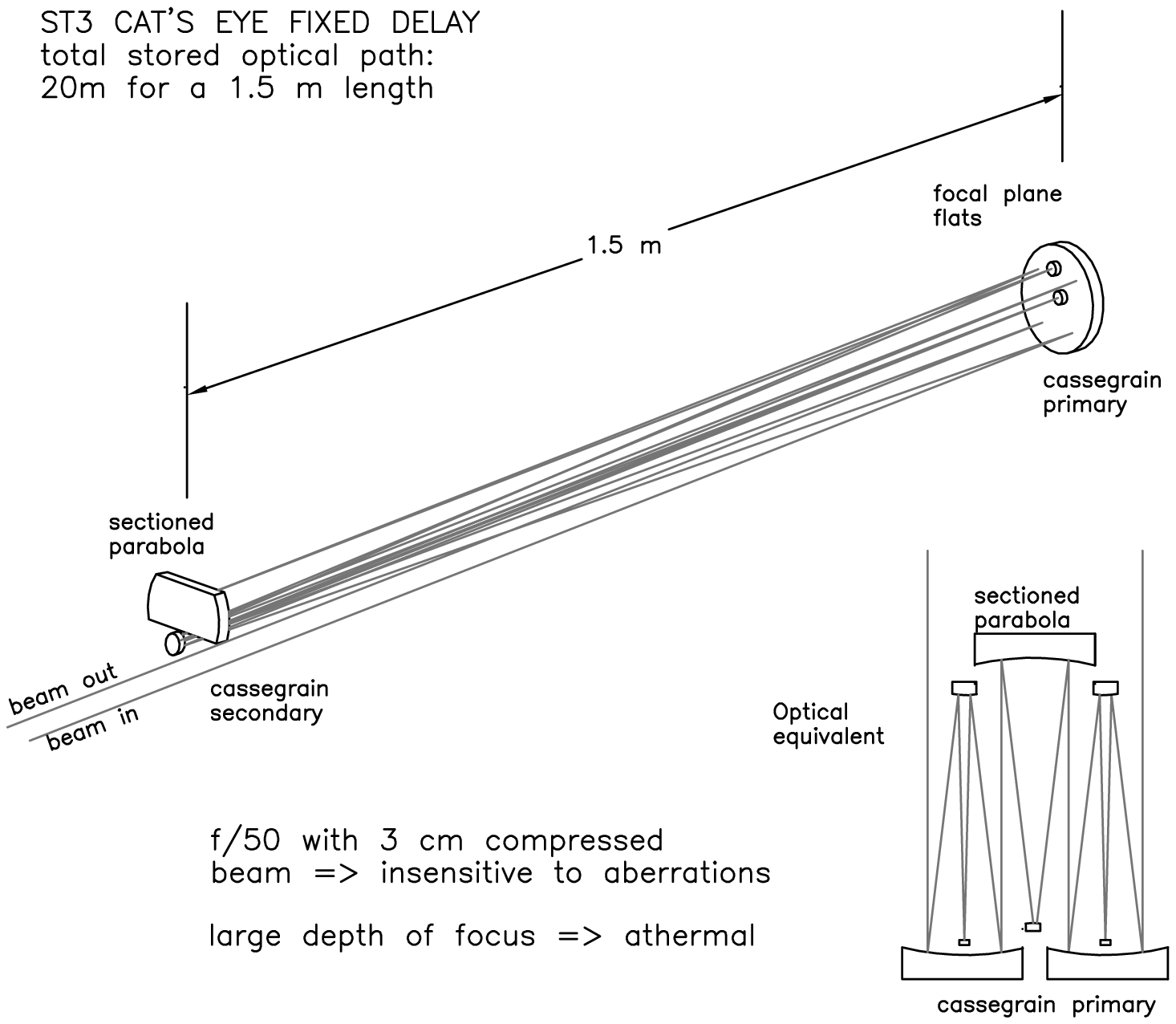,width=4.1in}
\caption{Fixed delay line perspective and schematic views.} \label{fig-2}
\end{figure}

Figure 3 indicates schematically the optical design for ST3 in the
adopted dual spacecraft configuration. The optical train is
almost completely planar throughout the system, and employs
an athermalized ultra-stable composite optical bench. In the
combiner spacecraft (which will function as a standalone 
fixed-baseline interferometer) a pair of outboard siderostats
feed into an afocal gregorian compressor with a 1 arcmin fieldstop at the internal focus. 

The 12 cm incoming beams are then compressed to 3 cm
and fed into the delay lines, one fixed and one movable. After 
this the beams enter the beam combiner. An outer 0.5 cm annular
portion of each beam is stripped off for guiding, and the
central 2 cm portion of the beam is used for fringe tracking (using
a single-element avalanche-photodiode detector in one of the combined
beams). The other combined 2 cm beam is dispersed in a prism and
integrated coherently on an 80 channel CCD fringe spectrometer.

\subsection{Fixed delay line}

Perspective and schematic views of the fixed delay line are shown
in Fig. 4. The design employs 3 nested cat's eye retroreflectors,
two of which are in a Cassegrain configuration and the third 
a Newtonian. As noted in the plot, the optics are very slow,
giving large depth of focus and minimal impact on wavefront 
distortion. Three of the 13 reflections  occur at foci and have
little wavefront effect. However, due to the large magnification
of the system, focal plane flats must be sized generously to
match field of view requirements. 

\acknowledgments
We thank M. Shao and M. Colavita for their invaluable suggestions
\& support of this work.
The research described here was carried out by the Jet Propulsion 
Laboratory, California Institute of Technology, under a contract
with the National Aeronautics and Space Administration.

\end{document}